\documentclass{article}

\usepackage{amsfonts}
\usepackage{amsmath}
\usepackage{amstext}
\usepackage{xspace}
\usepackage{amssymb}
\input psfig.sty

\newcommand{\N}{\mathbb N}

\newcommand{\E}{\mathbb E}

\begin{document}

\title{Toward a dynamical model for prime numbers} 

\author{Claudio Bonanno\footnote{Dipartimento di Matematica,
Universit\`a di Pisa, via Buonarroti 2/a, 56127 Pisa (Italy), email:
<bonanno@mail.dm.unipi.it>} \and Mirko S. Mega\footnote{Dipartimento
di Fisica, Universit\`a di Pisa, via Buonarroti 2, 56127 Pisa (Italy),
email: <mstefano@math.uc3m.es>}}

\date{}

\maketitle

\begin{abstract}
We show one possible dynamical approach to the study of the
distribution of prime numbers. Our approach is based on two complexity
methods, the Computable Information Content and the Entropy
Information Gain, looking for analogies between the prime numbers and
intermittency.
\end{abstract}

\section{Introduction} \label{sec:intro}

Nowadays one of the most famous open problem in mathematics is the
study of the distribution of prime numbers. This difficult problem has
been studied in the last two centuries by many mathematicians, using
many different techniques. More recently also in the physical
literature it is possible to find many papers devoted to prime
numbers, with two different points of view: a numerical approach using
methods developed for physical systems; a theoretical approach, where
prime numbers are found in some properties of physical systems.

This paper can be considered as a tentative to mix together the
mathematical and the physical approach. We analyze prime numbers using
numerical methods developed to study complexity in time series, and
look for interpretations of the results in the realm of dynamical
systems, in particular intermittent dynamical systems.

We first explain is Section \ref{sec:compl} the complexity methods we
use. Then in Section \ref{sec:inter_pn} we start to build a dynamical
approach to prime numbers, looking for a dynamical model, and testing
it by the complexity methods. It is interesting that the two methods
seem to show some differences, sign of how difficult is to give a
universal notion of complexity.

Finally in Section \ref{sec:distance} we turn back to mathematics, to
show that the prime numbers seem to have different statistical
properties with respect to our dynamical models. Hence both the
methods, that showed these discrepancies in different ways, seem to be
able, if used together, to make a deep analysis of time series.

\section{Complexity methods for time series} \label{sec:compl}

There are many methods used in literature to study the ``complexity''
of a time series. In this paper we used two methods that follow from
the interpretation of a time series as an information source:
Computable Information Content (\cite{argenti},\cite{licatone}) and
Diffusion Entropy (\cite{de1},\cite{de2}). Both methods aim to study the complexity
of an underlying dynamics that produce the time series. The two
methods have been studied also in \cite{allegrini}.

\subsection{Computable Information Content} \label{sec:inform}

Let $(X,T,\mu)$ be an invariant dynamical system, where $X$ is a
compact metric space (the phase space) with the Borel
$\sigma$-algebra, $T:X\to X$ is a measurable transformation of $X$,
and $\mu$ is a $T$-invariant measure on $X$, that is
$\mu(T^{-1}(A))=\mu(A)$ for all measurable sets $A\subseteq X$ (where
$T^{-1}(A)$ is the counter-image of the set $A$). 

One of the classical indicators of chaotic behavior of a dynamical
system is given by the {\it Kolmogorov-Sinai (KS) entropy} $h_\mu
(X,T)$, defined for probability $T$-invariant measures $\mu$ on $X$
(\cite{kolmog1},\cite{kolmog2}). Chaitin and Kolmogorov independently
introduced another method based on information theory to study chaotic
systems, the {\it Algorithmic Information Content (AIC)} (see
\cite{chaitin} and \cite{kolmogorov}).  The two indicators of chaos
are shown to coincide.

To study the information produced by a dynamical system, let $Z$ be a
finite partition of the phase space $X$ into measurable sets $\{
I_1,\dots,I_N\}$. To any point $x\in X$ we can associate an infinite
string $\omega =(\omega_0,\omega_1,\dots,\omega_n,\dots)$, where for
all $n\in \N$ it holds $\omega_n \in {\mathcal A}=\{1,\dots,N\}$ and
$T^n(x)\in I_{\omega_n}$. The set $\mathcal A$ is called the {\it
alphabet} associated to the partition $Z$, and the set of such strings
$\omega$ is denoted by $\Omega\subseteq {\mathcal A}^\N$. We call the map
$\varphi_Z :X \to \Omega$ the {\it symbolic representation} of $(X,T)$
relative to $Z$. The space $\Omega$ is a compact metric space, with
metric given by
$$d(\omega,\omega') = \sum_{n\in N} \
\frac{1-\delta(\omega_n,\omega'_n)}{2^{n}}$$ where $\delta(i,j)= 0$ if
$i\not= j$ and $\delta(i,j)=1$ if $i=j$. The Borel $\sigma$-algebra on
$\Omega$ is generated by the so-called {\it cylinders}
$C^{(k,h)}_\omega$, defined for all integers $k\le h$ and $\omega \in
\Omega$ by
$$C^{(k,h)}_\omega = \left\{ \omega'\in \Omega \ /\ \omega'_n =
\omega_n \ \forall \ n =k, \dots, h \right\}$$ Finally let
$\sigma:\Omega \to \Omega$ be the continuous transformation of
$\Omega$ given by $(\sigma(\omega))_n = \omega_{n+1}$ for all $n\in
\N$. The map $\sigma$ is called the {\it shift map}. Any $T$-invariant
measure $\mu$ on $X$ induces by the symbolic representation a
$\sigma$-invariant measure $\nu:=(\varphi_Z)_* \mu$ on $\Omega$. Then
$(\Omega, \sigma, \nu)$ is a dynamical system, called the {\it
symbolic dynamical system}.

When the symbolic representation $\varphi_Z$ is injective then the
dynamical systems $(X,T,\mu)$ and $(\Omega,\sigma,\nu)$ are
topologically conjugate. In this case the entropies $h_\mu (X,T,Z)$
and $h_\nu (\Omega,\sigma)$ coincide, for any probability measures
$\mu$ and $\nu=(\varphi_Z)_* \mu$, respectively $T$ and
$\sigma$-invariant. Here $h_\mu(X,T,Z)$ is the KS entropy of the
dynamical system $(X,T,\mu)$ with respect to the partition $Z$. It
holds $h_\mu(X,T) = \sup \ \{ h_\mu (X,T,Z) \ /\ Z \hbox{ is a finite
partition of } X \}$ and the supremum is attained on {\it generating}
partitions.

The method of symbolic dynamics tells us that to study the information
produced by the dynamical system $(X,T,\mu)$, we can look at the
different symbolic representations of the system, and in particular to
those corresponding to generating partitions. Then we need to study
the information in some sense contained in strings. One way is given
by the {\it Algorithmic Information Content (AIC)}.

Let $s$ be a finite string on an alphabet $\mathcal A$, and let $C$ be a
universal Turing machine, then if $p$ denotes a program given to $C$
as input, we define
\begin{equation} 
AIC(s) = \min \ \left\{ |p| \ /\ C(p)=s \right\}
\label{def:aic}
\end{equation}
where $|p|$ is the binary length of the program $p$. In words, the
$AIC$ of a finite string $s$ is the binary length of the smallest
program $p$ that gives $s$ as output when run on a universal Turing
machine.

The idea underlying the definition of the KS entropy is to estimate
the rate with which the information necessary to reconstruct an orbit
of the dynamical system increases with time. So, analogously, we study
the rate with which the information contained in an infinite string
increases with the length of the string. Formally, let $\omega \in
\Omega$ be an infinite string, then the {\it complexity} $K(\omega)$
is defined as
\begin{equation}
K(\omega) = \limsup_{n\to \infty} \ \frac{AIC(\omega^n)}{n}
\label{def:compl}
\end{equation}
where $\omega^n = (\omega_0,\dots,\omega_{n-1})$ is the $n$-long
string made by the first $n$ symbols of the infinite string $\omega$.

At this point it is immediate, by the method of symbolic
representation of a dynamical system, to introduce a notion of
complexity for orbits of a dynamical system using the AIC. Let indeed
$(X,T,\mu)$ be an invariant dynamical system, and let $Z$ be a finite
measurable partition of $X$. Then via the map $\varphi_Z$ to each
point $x\in X$ we associate an infinite string $\omega \in \Omega
\subseteq {\mathcal A}^\N$, where $\mathcal A$ is the finite alphabet
associated to $Z$. Then the {\it complexity $K(x,Z)$ of a point $x$
with respect to the partition $Z$} is given by
\begin{equation}
K(x,Z) := K(\varphi_Z(x)) = \limsup_{n\to \infty} \ \frac{AIC(x,n,Z)}{n}
\label{def:complpunti}
\end{equation}
where $AIC(x,n,Z) := AIC(\varphi_Z(x)^n)$.

The complexity of a point $x$ and the KS entropy of a dynamical
system are related by the following result (see
\cite{brudno},\cite{licatone}): let $(X,T,\mu)$ be an invariant
dynamical system, and let $\mu$ be a probability measure, then
\begin{equation}
\int_X  K(x,Z) \ d \mu = h_\mu (X,T,Z)
\label{teo:lica}
\end{equation}
for any finite measurable partition $Z$ of $X$. In particular if the
probability measure $\mu$ is also ergodic then 
\begin{equation}
K(x,Z)= h_\mu (X,T,Z)
\label{teo:brud}
\end{equation}
for $\mu$-almost any $x\in X$. To obtain the KS entropy of a system,
one can choose a generating partition $Z$ and
$h_\mu(X,T,Z)=h_\mu(X,T)$. If the invariant ergodic measure $\mu$ is
infinite then (see \cite{tesi} for exact hypothesis on systems with
infinite measures)
\begin{equation}
K(x,Z)=0
\label{teo:tesi}
\end{equation}
for $\mu$-almost any $x\in X$.

This result on complexity is not particular relevant for the case of
probability invariant measures when $h_\mu(X,T)=0$, but still the
system is chaotic (these systems are called {\it weakly chaotic}), or
for infinite invariant measures. In these cases what one can do is to
look at the rate of increase of the $AIC$ for the orbits of the
system. Indeed, if $h_\mu(X,T,Z)>0$ then, for a finite ergodic system,
it holds $AIC(x,n,Z)\sim h_\mu(X,T,Z) n$ for $\mu$-almost any $x$, but
for ergodic weakly chaotic systems and for ergodic infinite systems it
holds $AIC(x,n,Z) = o(n)$. Hence what one can do to classify the
complexity of the system is to study the asymptotic behavior of the
$AIC$. This has been done for some examples of weakly chaotic and
infinite systems, in \cite{licatone} the results are collected.

The problem with this approach is that the Algorithmic Information
Content is a function that is not computable, in the sense that does
not exist an algorithm able to compute the AIC for any finite
string. This result is related to the Turing machine halting problem.
Hence what one can do in practice is to study an algorithm which
approximates the AIC. We used {\it compression algorithms}, that are
formally defined in \cite{licatone}. Given a finite string $s\in
{\mathcal A}^*$, the set of all finite words in the alphabet $\mathcal A$,
a compression algorithm takes it as input and gives as output a finite
string $s' \in \{0,1\}^*$, that is in some sense a string which
contains all the information necessary to reconstruct $s$. We use as
approximation of $AIC(s)$ the length of the compressed string $s'$. We
call it {\it Computable Information Content (CIC)} of $s$. In the
following we use a particular compression algorithm called {\it
CASToRe} (introduced in \cite{licatone}), that has been tested on some
weakly chaotic and infinite systems, giving good approximations of the
AIC. So we have $CIC(s) = |CASToRe(s)|$, where $|\cdot|$ denotes the
length of a string.

\subsection{Diffusion Entropy (DE)} \label{sec:gain}

The main idea of this method is to investigate the time series
complexity building up a diffusion process, without any kind of
pre-processing of the series. This technique is based principally on
the Continuous time random walk (CTRW) and on the Generalized Central
Limit Theorem(GCLT). It is very sensitive to time randomness, when the
deviation from Poisson statistics generates L\'{e}vy diffusion, and to
the trend of a non-stationary series, as in prime numbers. Let
$\xi_i$, with $i=1,ŽÂ…., M$, be a sequence of $M$
numbers. The purpose of the DE is to establish the possible existence
of a scaling, either normal or anomalous, in the most efficient way as
possible, without altering the data with any form of detrending. Let
$l$ be a integer number, fitting the condition $1\le l \le M$. This
integer number will be referred as time. For any given $l$, there are
$M-l+1$ sub-sequence defined by

\begin{equation} \label{uno}
\xi_i^{(s)}\equiv \xi_{i+s} \quad \quad s=0,...,M-l 
\end{equation}

For any of this sub-sequence, a diffusion trajectory is built up,
labeled with the index $s$, defined by 
\begin{equation} \label{due}
x^{(s)}(l)=\sum_{i=1}^l \xi_i^{(s)}
\end{equation}

This position can be imagined as referring to a Brownian particle
that, at regular intervals of time, has been jumping forward or
backward, according to the prescription of the corresponding
sub-sequence. This means that the particle, before reaching the
position that it holds at time $l$, has been making $l$ jumps. The
jump made at the $i$-th step has the intensity $|\xi_i^{(s)}|$ and is
forward or backward according to whether the number $\xi_i^{(s)}$ is
positive or negative.  Now the entropy of this diffusion process can
be evaluated. To do that, the $x$-axis is partitioned into cells of
size $\epsilon(l)$. When this partition is made, the cells are
labeled.  The number of how many particles are found in the same cell
at a given time $l$ is denoted by $N(l)$. This number is used to
determine the probability that a particle can be found in the $i$-th
cell at time $l$, $p_i(l)$, by means of
\begin{equation}
p_i(l) \equiv \frac{N_i(l)}{(M-l+1)} 
\end{equation}
At this stage, the entropy of the diffusion process at time $l$ is
determined, and reads
\begin{equation}
S_{\epsilon(l)}(l)= - \sum_i\ p_i(l) \log [p_i(l)] 
\end{equation}
 
The easiest way to proceed with the choice of the cell size,
$\epsilon(l)$, is to assume it independent of $l$ and determine it by
a suitable fraction of the square root of the variance of the
fluctuation in the data $\xi_i$.  A little comment on the way used to
define the trajectories: the method is based on the idea of a moving
window of size $l$ that makes the $s$-th trajectory closely correlated
to the next. The two trajectories have $l-1$ values in common. A
motivation of this choice is a possible connection with the
Kolmogorov-Sinai (KS) entropy. The KS entropy of a symbolic sequence
is evaluated by moving a window of size $l$ along the sequence. Any
window position corresponds to a given combination of symbols, and
from the frequency of each combination, it is possible to derive the
Shannon entropy $S(l)$.  The KS entropy in given by the limit
$\lim_{l\to \infty}S(l)/l$. Hence the idea is that the same sequence
analyzed with the DE method, should yield at large values of $l$ a
well-defined scaling $\delta$. As a simplifying assumption let
consider large values of times to make the continuous assumption
valid. In this case, the trajectories built up with the above
illustrated procedure correspond to the following equation of motion:
\begin{equation}
\frac {dx}{dt}=\xi(t)
\end{equation}
where $\xi(t)$ denotes the value of the time series under study
at time $t$. This means that the function $\xi(l)$ is
depicted as a function of $t$, thought of as a continuous time $t=l$. 
In this case, the Shannon entropy reads 
\begin{equation} \label{12}
S(x,t)=-\int_{-\infty}^{\infty} \ p(x,t) \log [p(x,t)] dx
\end{equation}
being $p(x,t)$ the probability distribution function of the $x$
variable at time $t$.

It has to be pointed out that scaling is a property implying that 
the probability distribution function can be read as
\begin{equation} \label{13}
p(x,t)=\frac{1}{t^\delta}F\left(\frac{x}{t^\delta}\right)
\end{equation}  
Plugging equation (\ref{12}) into equation (\ref{13}), and after a
little bit of algebra, $S(t)$ reads
\begin{equation}
S(x,t)= A + \delta \log (t) 
\end{equation}
where
\begin{equation}
A=\int_{-\infty}^{\infty}\ F(y)\log [F(y)] dy 
\end{equation}
It is now evident that this kind of techniques to detect scaling does
not imply any form of detrending.  This technique only works correctly
in the case of a stationary time series, on the contrary, as can be
analytically and numerically proved, if there is a trend we find a
scaling with $\delta=1$, as in the case of the prime numbers series.
Whereas if the data $\xi_i$ are the output of a process of independent
and identically distributed random variables, then by the Central
Limit Theorem the probability distribution function $p(x,t)$ converges
to the Gauss distribution as $t\to \infty$, hence we find a scaling
parameter $\delta=0.5$.

On the same idea of not pre-processing the real data, a technique of
conditional entropy, called {\it Entropy Information Gain (EIG)}
\cite{EIG} can be used, joined with the DE method, to detect the
residual entropy associated to the independent fluctuations overlapped
to the original trend. Let $\xi_i$ be the original data and
$\chi_i$ the time series produced using the hypothetical trend, then the corresponding
trajectories are :
\begin{equation}
x^{(s)}(l)=\sum_{i=1}^{l} \xi_i^{(s)} \quad \mbox{and} \quad y^{(s)}(l)=\sum_{i=1}^{l} \chi_i^{(s)} 
\end{equation}
where we followed the same notations of equations (\ref{uno}) and
(\ref{due}). At this point we can repeat the same argument that we did
for the DE to use continuous time, and define the EIG of the original
data with respect to the hypothetical trend as the function $I(x,y,t)$ given by
\begin{equation}
I(x,y,t)=-\int_{-\infty}^{+\infty} p(y,t)dy\int_{-\infty}^{+\infty} p(x|y,t) \log p(x|y,t) dx 
\end{equation}
where $p(x,t)$ and $p(y,t)$ are the probability distribution functions
associated to the trajectories $x$ and $y$ at time $t$, and
$p(x|y,t)=p((x,y),t)/p(y,t)$, being $p((x,y),t)$ the joint probability
distribution function. After some simple algebra we find (see equation (\ref{12}))
\begin{equation}
I(x,y,t)=S((x,y),t)-S(y,t)
\end{equation}
that is an additive relation. Then it is evident that the better the
hypothetical trend approximates the real time series, the smaller is
the EIG. Moreover, applying the method of DE to the EIG, looking for
scaling properties for the function $I(x,y,t)$, we interpret a
$\delta=0.5$ as a result of independent fluctuations in the original
data, overlapped to the hypothetical trend.

\section{Intermittency and the prime numbers} \label{sec:inter_pn}

We now apply our methods to two different time series, studying the
differences and the analogies. We first study the time series coming
from an {\it intermittent} dynamical system, the {\it Manneville
map}. The Manneville map is a dynamical system defined on the interval
$I=[0,1]$ by
\begin{equation}
T_z(x)= x + x^z \ (\hbox{mod } 1) \qquad z>1
\label{def:mann}
\end{equation}
where $z$ is a real parameter. This map has the feature to have a
fixed point in $x=0$ which is not hyperbolic, indeed $T'_z(0)=1$ for
all $z>1$. This makes the system interesting study, in particular for
values of the parameter $z\ge 2$. Indeed in that range the invariant
measure that is absolutely continuous with respect to the Lebesgue
measure is infinite. For further properties we refer to
\cite{allegrini},\cite{boga}.

This map was originally introduced in \cite{mann} as a simple
dynamical model for turbulence. Indeed the presence of the
non-hyperbolic fixed point at the origin generates a phenomenon known
as {\it intermittency}: an alternation of long laminar phases and
short chaotic bursts. This intermittency is generated by the fact that
when a point goes close to the origin it takes a long time to
``escape'' from it (the laminar phases), then it has a short period of
time when it remains far from the origin (the chaotic burst), but then
it falls again near the origin, and this process repeats on and on.

The behavior of the AIC for orbits of the Manneville map is studied
in \cite{boga}. It is proved that,
\begin{equation}
\E[AIC(x,n,Z)] \sim \left\{
\begin{array}{cl}
n & z<2 \\[0.2cm]
\frac{n}{log n} & z=2 \\[0.2cm]
n^{\frac{1}{z-1}} & z>2
\end{array}
\right.
\label{aicmann}
\end{equation}
where $\E[\cdot]$ denotes the mean with respect to the Lebesgue
measure and it has been chosen the partition $Z=\{ [0,\tilde x),
[\tilde x, 1] \}$, being $\tilde x$ the real number in $(0,1)$ such
that $x+x^z =1$. The same behavior is obtained for the Computable
Information Content $\E[CIC(x,n,Z)]$ using $CASToRe$
(\cite{licatone}).

\vskip 0.5cm
The second time series we study is obtained from the series of prime
numbers. Let $\{ 1,2,3,\dots \}$ be the integer numbers, we associate
to each number the symbol ``1'' if it is a prime number, and the
symbol ``0'' if it is not. So we construct an infinite string $\bar
\omega \in \{ 0,1\}^\N$, whose first symbols are $\bar \omega=
(0,1,1,0,1,0,1,\dots)$. The most famous question on prime numbers is
``how many prime numbers are there in the first $n$ integers?''. This
question becomes in our formulation ``how many ``1''s are there in the
first $n$ symbols of $\bar \omega$?''. We refer to \cite{apostol} for
the formal approach. The first answer to this question was given by
Gauss in 1849, and his result is contained in the
\vskip 0.3cm
\noindent {\bf Prime Numbers Theorem.} If $\pi(n)$ is the number of
primes up to the first $n$ integers, it holds
$$\pi(n) \sim \mbox{Li}(n) := \int_2^n \ \frac{dt}{\log t} \ \sim
\frac{n}{\log n}$$
\vskip 0.3cm 

We now study what happens if we want to study the information content
of the string $\bar \omega$. Since there are many algorithms to
generate the prime numbers (think for example to Eratosthenes' sieve),
if we want to estimate $AIC(\bar \omega^n)$ it is clear that it is of
the order $\log n$, since the algorithm that generates the primes only
needs to know at what integer $n$ it has to stop (this is done with
$\log n$ bits of information). So the shortest program has length of
the order $\log n$. But we imagine to forget the origin of the string
$\bar \omega$ and look for its information content. 

In \cite{boga} it is shown that the compression of a binary string
done simply remembering how many ``0''s there are between two
consecutive ``1''s, is the ``best'' compression for strings of the
Manneville map (``best'' in the sense that it approximates the
behavior of the AIC up to a constant). This fact is due to the
property of strings from the Manneville map that there are long
sequences made only of ``0''s (correspondent to the laminar phases)
and short sequences with chaotic alternation of ``0''s and ``1''s (the
chaotic bursts), and to the property that the passage through a
chaotic burst can be considered as a phenomenon deleting the memory of
the system, so what happens after a symbol ``1'' does not depend on
what happened before. The string $\bar \omega$ of prime numbers has
the same property of long sequences of ``0'' and rare appearances of
``1''s, so we could try to apply this kind of compression to $\bar
\omega$ to see what happens, even if the property of loss of memory is
false for prime numbers. It is not difficult to see that the length of
the new string obtained is related to the number of symbols ``1'' in
the original string. So it is related to the function $\pi(n)$. This
argument shows that we could imagine the string $\bar \omega$ as given
from one of the orbits of the Manneville map with $z=2$, both having
the same asymptotic behavior of the information content (see equation
(\ref{aicmann})).

We remark that this approach to study prime numbers is similar to the
probabilistic approach introduced by Cram\'er in \cite{cramer1}, that
is we assume that the string $\bar \omega$ is one of a family of
strings, on which there is a probability measure $\mu$. If then a
property holds for $\mu$-almost all strings of the family, then there
is a ``good'' chance that it holds also for $\bar \omega$. Of course
it is not formally true, the string $\bar \omega$ could be in the set
of null $\mu$-measure for which the property does not hold, but
certainly it is a good first guess for the properties of $\bar
\omega$. Our aim is to look for a dynamical system $(X,T)$ whose
orbits have (in some probabilistic sense) the same statistical
properties of $\bar \omega$, so that we can think of $\bar \omega$ as
the symbolic representation of an orbit of the system, and guess some
new properties of $\bar \omega$ from those of the system.

Following the idea explained above, we compare the statistical
properties of strings generated by the Manneville map $T_2$ and the
infinite string $\bar \omega$. We apply the methods introduced in
Section \ref{sec:compl} to study the complexity of the time series.

\begin{figure}[ht]
\psfig{figure=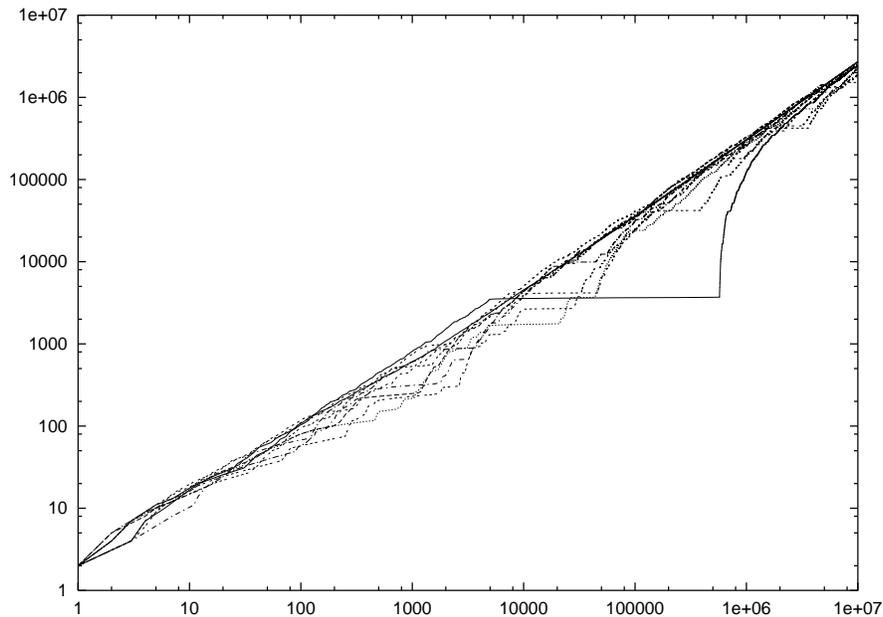,width=12cm,angle=270}
\caption{\it The information content measured using $CASToRe$ for the
series of the prime numbers and ten different orbits of the Manneville
map $T_2$. The scale is bi-logarithmic.}
\label{bilog_primi_mann2}
\end{figure}

We first apply the compression algorithm $CASToRe$. We built the first
$10^7$ symbols of the string $\bar \omega$ and $10$ orbits of the
Manneville map $T_2$ of the same length, corresponding to different
initial conditions (this because we remark that for the AIC of $T_2$
we only have results in mean with respect to the Lebesgue measure). In
Figure \ref{bilog_primi_mann2} it is shown the behavior of the
information content for all the series. 

\vskip0.3cm
\vskip 0.3cm

The graph is in bi-logarithmic scale, hence the behavior $CIC(n) \sim
\frac{n}{\log n}$ becomes a straight line with angular coefficient
$1$, perturbed by a logarithm, that is smaller than $7 \log 10$. In
this graph it is difficult to distinguish the different curves, but
there is one that is practically straight, the one of $\bar
\omega$. Of course the algorithm $CASToRe$ is only an approximation of
the AIC of our series, hence we cannot expect the curves to be exactly
of the order $\frac{n}{\log n}$, but they are approximations to
it. However our aim is not to estimate the information content of the
series, but simply to make a comparison between their behaviors, hence
we can neglect the effects due to the use of the particular
compression algorithm since they are the same on each series.

\begin{figure}[ht]
\psfig{figure=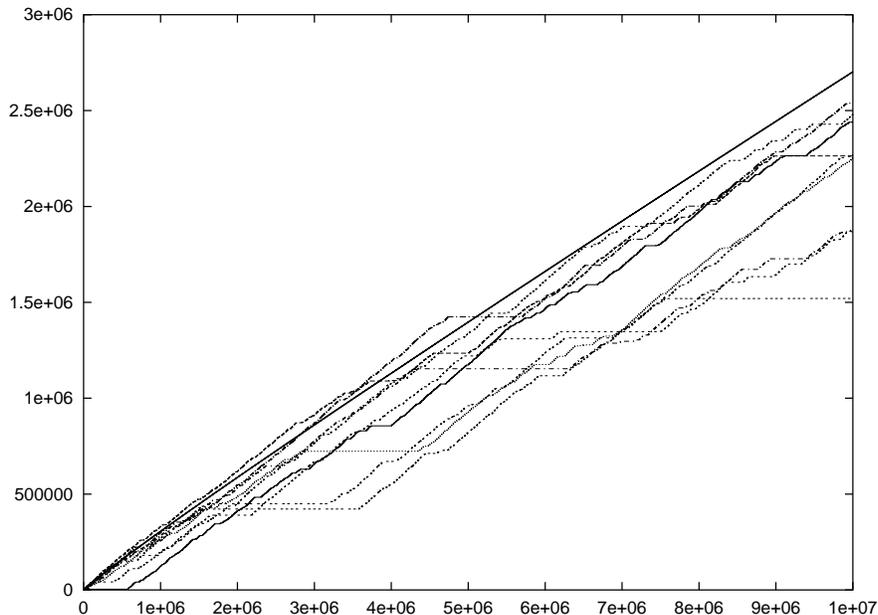,width=12cm,angle=270}
\caption{\it The same as in Figure \ref{bilog_primi_mann2}, but in
bi-linear scale.}
\label{primi_mann2}
\end{figure}

From this first figure we could then conclude that the Manneville map
$T_2$ is a ``good'' dynamical system to generate the series of the
prime numbers, in the sense explained above. However if we look at the
same graphs in bi-linear scale (Figure \ref{primi_mann2}), we notice a
small difference between the straight line (the information content of
$\bar \omega$) and the other curves. 

\vskip0.3cm
\vskip0.3cm

This small difference is not visible in the bi-logarithmic figure,
since the bi-logarithmic scale reduces the differences at high values
of $n$. Hence the information content of the string $\bar \omega$
seems to be asymptotically slightly different from the mean
information content of orbits of $T_2$.

This difference can be explained in terms of the approximation done in
the Prime Numbers Theorem. Indeed the function Li$(n)$ can be estimated
using the method of integration by parts, obtaining the approximation
$$\mbox{Li}(n) \sim \frac{n}{\log n} + \frac{n}{\log^2 n} + {\mathcal
O}\left( \frac{n}{\log^3 n} \right)$$ and this shows that the number
of primes in the first $n$ integers (that is the number of symbols
``1'' in the first $n$ symbols of the string $\bar \omega$) grows
slightly faster than $\frac{n}{\log n}$.

\begin{figure}[ht]
\psfig{figure=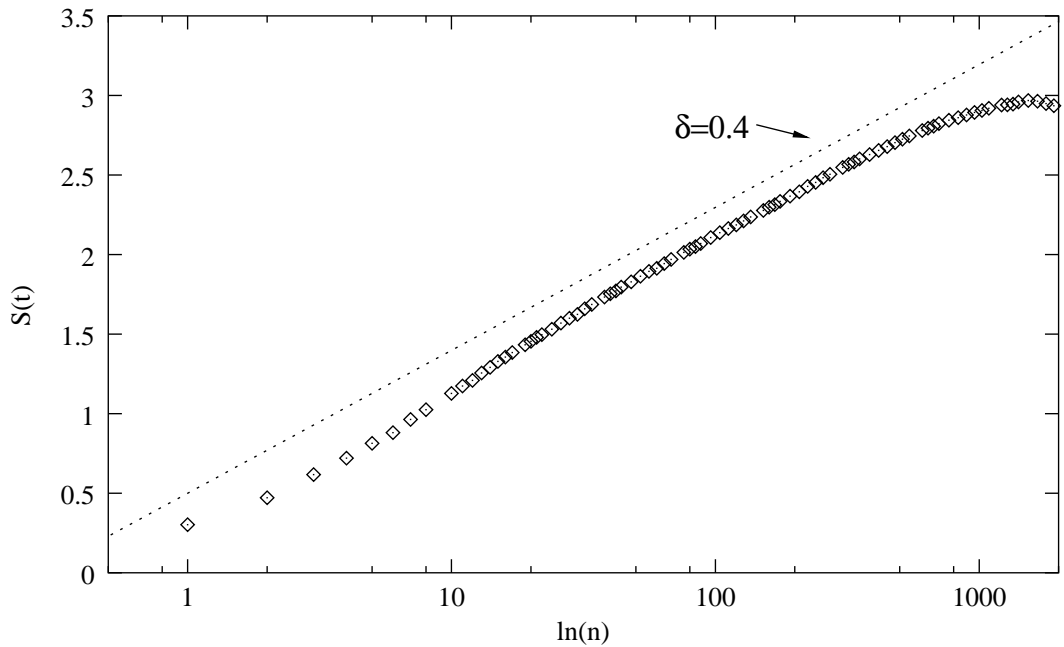,width=14cm,angle=0} 
\caption{\it EIG for the prime numbers series, using as trend the
first approximation of the function Li$(n)$.}
\label{primivsnsulog}
\end{figure}

Then we apply the Entropy Information Gain method to the same series.
First of all we study how good is the approximation of the function
$\pi(n)$ given in the Prime Numbers Theorem. We build up the different
trajectories to analyze just by the sum of the number of primes
contained in the strings of length $l$. We use for the hypothetical
trend the first approximation of the Prime Numbers Theorem, $\pi(n)
\sim \frac{n}{\log n}$. In Figure \ref{primivsnsulog} it is shown the
behavior of the residual Shannon entropy, that follows a straight
line with angular coefficient $\approx 0.4$. 

\vskip0.3cm
\vskip0.3cm

\begin{figure}[ht]
\psfig{figure=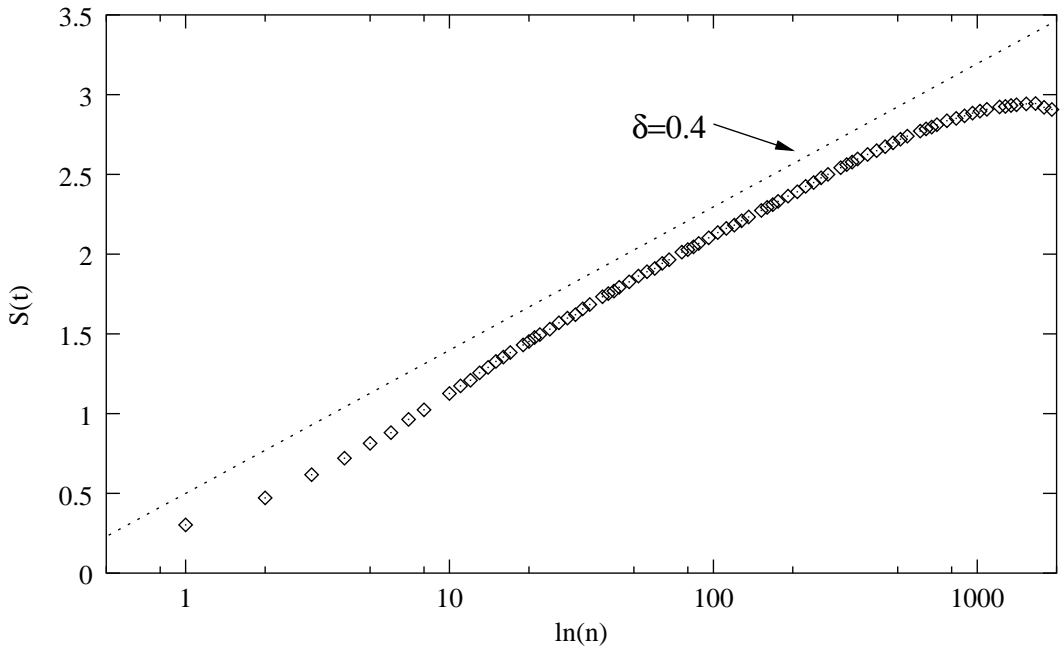,width=14cm,angle=0}
\caption{\it The same as in Fig \ref{primivsnsulog}
but with the first two orders of approximation for Li$(n)$.}
\label{primivs2ord}
\end{figure}

This value of the $\delta$, close to the Gaussian value $0.5$, seems
to suggest that all the information of the primes series is contained
in the Prime Number Theorem. To better investigate about the primes
trend, we have also calculated the EIG with more terms approximating
the function Li$(n)$, Figure \ref{primivs2ord}, but the results
doesn't change. This equality could be a limit for the sensibility of
this method, that seems to feel only the first approximation of a
hypothetical trend.  

\vskip0.3cm
\vskip0.5cm

\begin{figure}[ht]
\psfig{figure=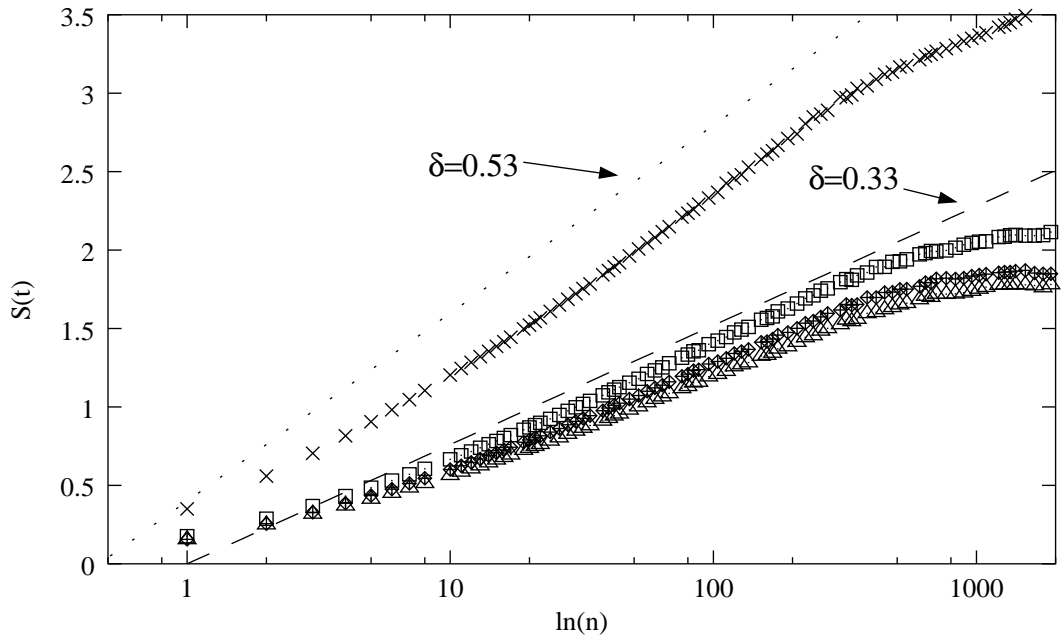,width=14cm,angle=0}
\caption{\it EIG between the prime numbers and the different
Manneville maps. The $\delta = 0.53$ correspond to the $T_2$
map, while the $\delta=0.33$ to the new hybrid maps.}
\label{EGmanntot} 
\end{figure}

At this point we have calculated the EIG using as data the prime
numbers series and as hypothetical trend the series of orbits produced
by the Manneville Map for $z=2$, Figure \ref{EGmanntot}. Also in this
case, the angular coefficient of the asymptotic straight line is
$\approx 0.5$, due to the fact that in first approximation, this
Manneville map contains the same information of the Prime Numbers
Theorem, even if the compression method has shown some differences,
that if coming from the approximation of the function Li$(n)$ are
stationary, so cannot be interpreted as independent fluctuations on
the trend.

\vskip 0.1cm

So far we have applied theoretical results from number theory and from
the theory of information content in intermittent dynamical systems,
obtaining a ``close'' relationship between the prime number series
$\bar \omega$ and the orbits of the Manneville map $T_2$. To find a
better dynamical model for the prime numbers, in the sense explained
above, we could slightly perturb the map $T_2$ and see what happens
for the information content. This is what is done below based only on
numerical experiments done using the compression algorithm $CASToRe$.

The statistical properties of the maps $T_z$ are induced by the
behavior of the maps near the origin (the non-hyperbolic fixed
point). Hence we have to perturb $T_2$ changing its behavior near the
origin. We have chosen the following family of maps: given two
positive real numbers $a,b$, the maps $T_{(a,b)}$ are defined by
\begin{equation}
T_{(a,b)}(x) = x + x^2 + a x^b \qquad x\in (0,\tilde x)
\label{def:nuove_mann}
\end{equation}
where $\tilde x$ satisfies $\tilde x + \tilde x^2 + a \tilde x^b =1$,
and $T_{(a,b)}(x)$ is randomly chosen in $(0,1)$ according to the
Lebesgue measure when $x>\tilde x$.

\begin{figure}[ht]
\psfig{figure=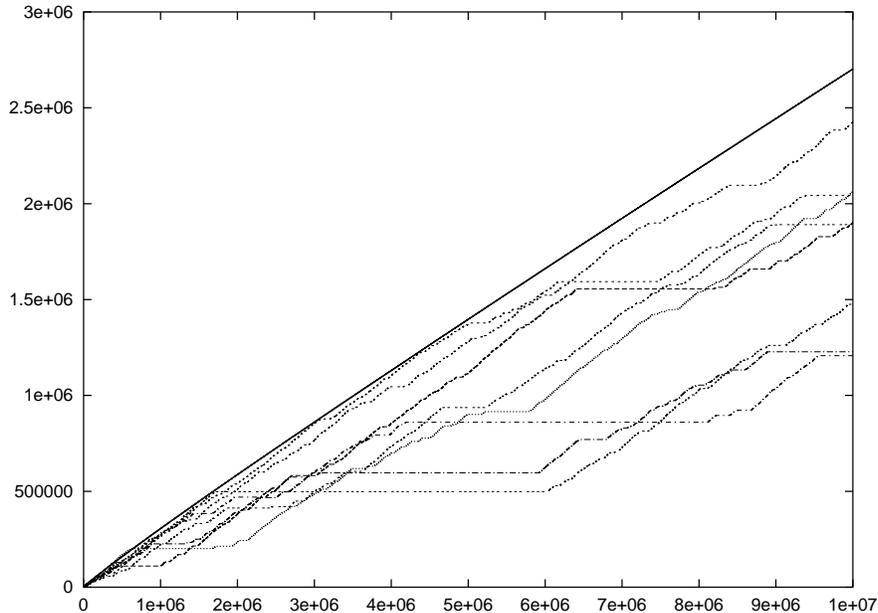,width=12cm,angle=270}
\caption{\it The information content measured using $CASToRe$ for the
series of the prime numbers and ten different orbits of the map
$T_{(1,3)}$. The scale is bi-linear.}
\label{primi_new3}
\end{figure}

\begin{figure}[ht]
\psfig{figure=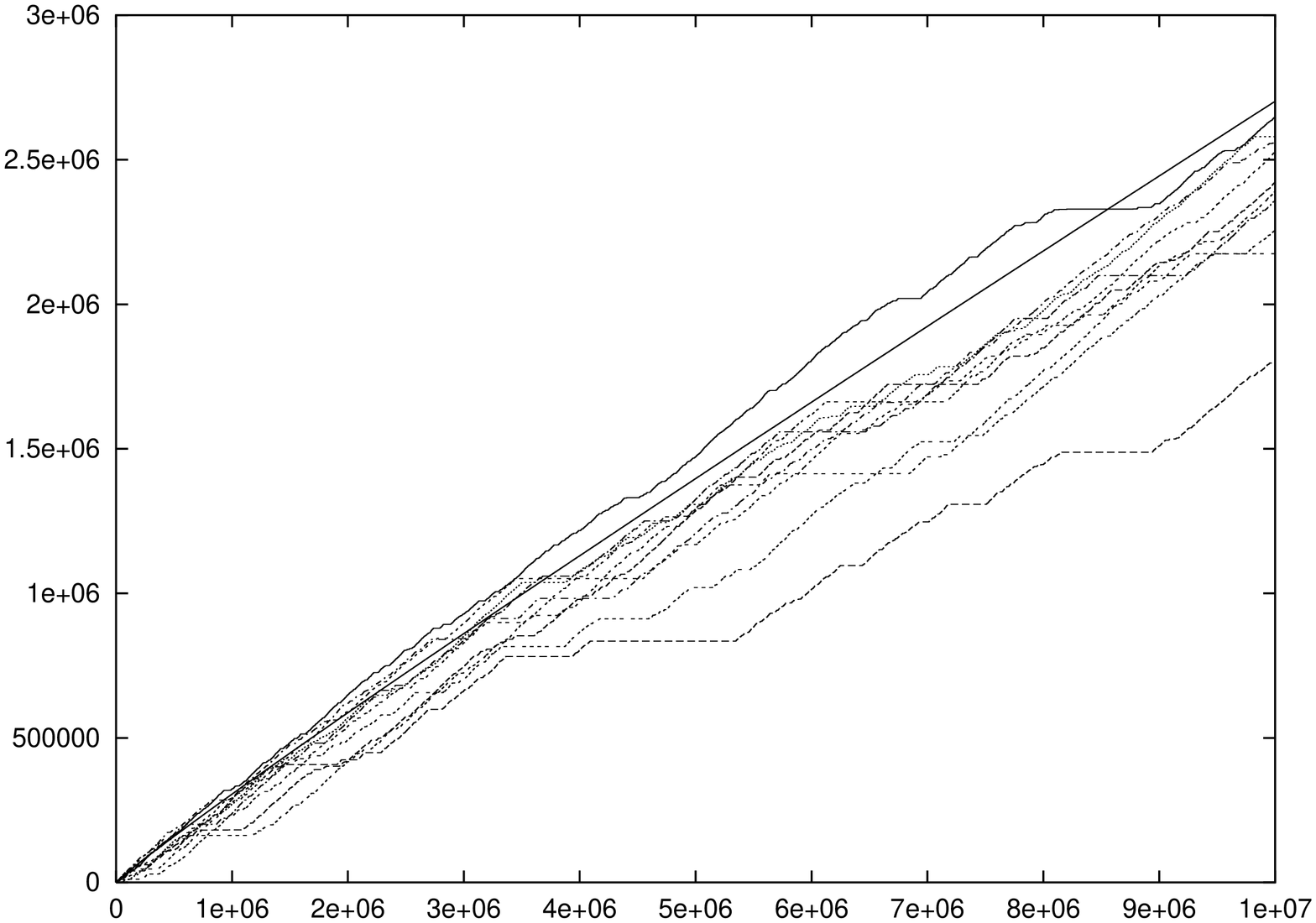,width=13cm,angle=270}
\caption{\it The same as in Figure \ref{primi_new3}, but for the map
$T_{(0.1,2)}$.} 
\label{primi_new0.1x2}
\end{figure}

\begin{figure}[ht]
\psfig{figure=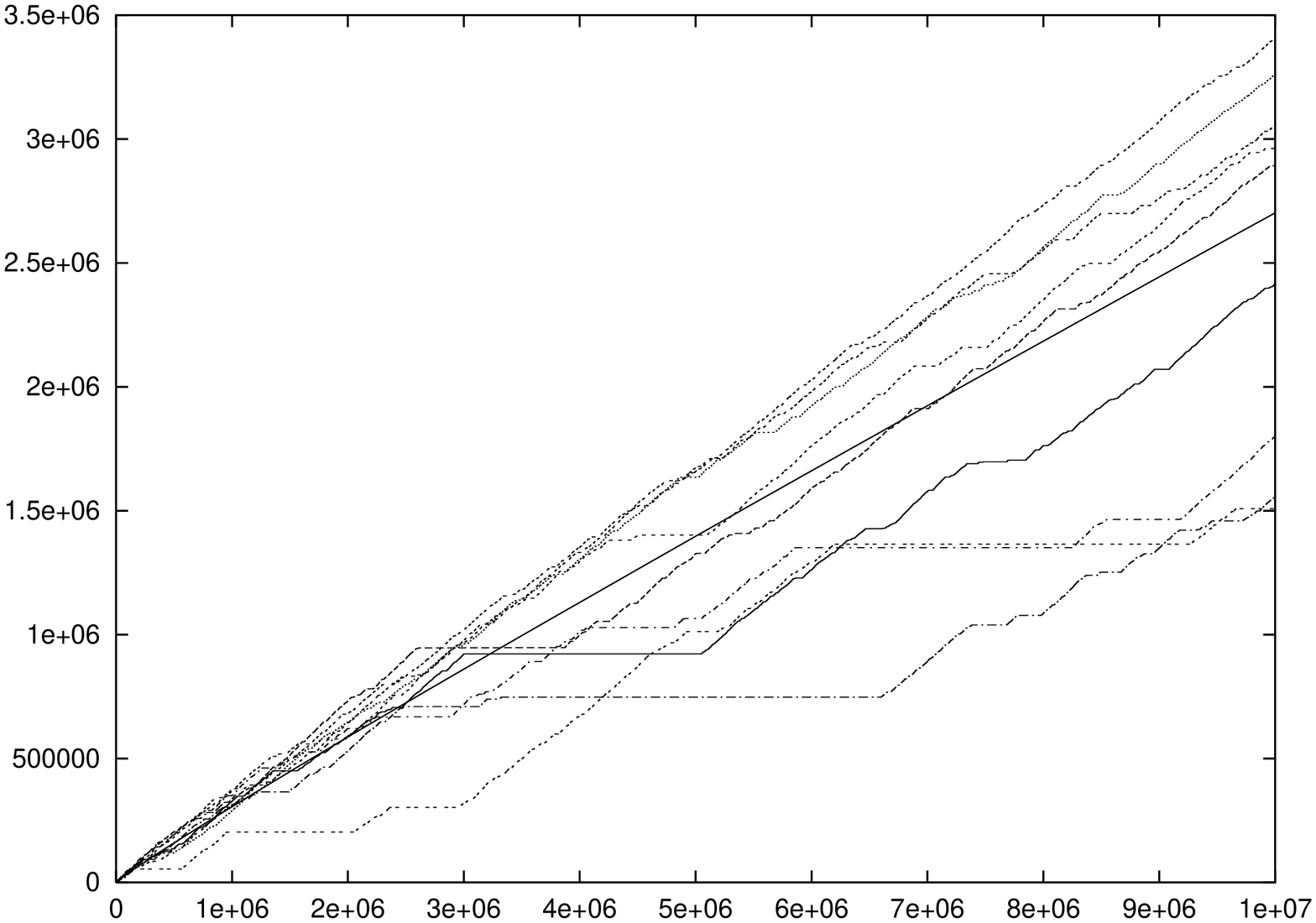,width=13cm,angle=270}
\caption{\it The same as in Figure \ref{primi_new3}, but for the map
$T_{(0.5,2)}$.} 
\label{primi_new0.5x2}
\end{figure}

\begin{figure}[ht]
\psfig{figure=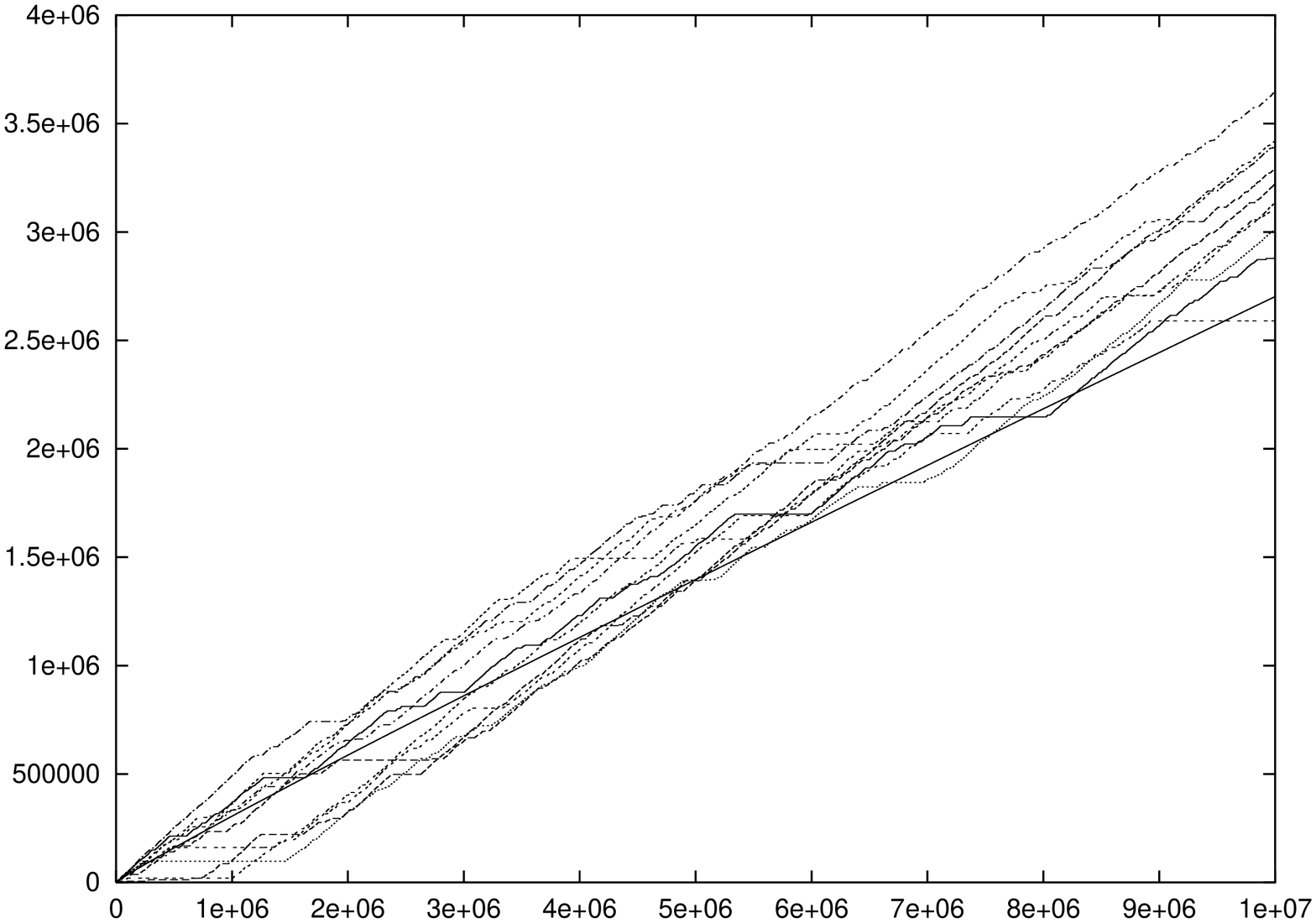,width=13cm,angle=270}
\caption{\it The same as in Figure \ref{primi_new3}, but for the map
$T_{(1,2)}$.} 
\label{primi_new1x2}
\end{figure}

In Figures \ref{primi_new3}-\ref{primi_new1x2} the results of the
experiments are shown. Each figure is in bi-linear scale, and there
are plotted the curves corresponding to the information content of the
string $\bar \omega$ (the straight line) and of some different orbits
of the maps $T_{(a,b)}$, for $(a,b)$ equal respectively to $(1,3)$,
$(0.1,2)$, $(0.5,2)$ and $(1,2)$. From the figures, we obtain as a
``good'' guess for a dynamical model of the series of prime numbers
the map $T_{(0.5,2)}$, even if in this case the results show a big
difference for the behavior of the information content of different
orbits (remember we have only results in mean for intermittent maps
with an infinite invariant measure).

\vskip0.3cm

\vskip0.3cm

On the same line, we have calculated the EIG between the prime numbers
series and the different perturbed Manneville maps $T_{(a,b)}$. The
results are shown in Figure \ref{EGmanntot}: in this figure there is
the EIG result for $T_2$, with $\delta = 0.53$, whereas all the
perturbed maps give a $\delta=0.33$, sign of an anti-correlation
behavior for the residual Shannon entropy. Hence it seems that now
the perturbation to the Manneville map gives better results from the
point of view of compression algorithms, but the results seem worse
from the point of view of the EIG.

\section{Waiting times and distance between primes} \label{sec:distance}

In the analysis of the statistical behavior of time series, an
important feature is the distance between two different events. In the
orbits of intermittent dynamical systems the events are the chaotic
bursts. The importance of these events for the intermittent systems is
reflected by the fact that the ``best'' compression was done by simply
recording how much time an orbit spends in the laminar zone (how many
symbols ``0'' there are in the series) between two different chaotic
bursts (corresponding to the appearance of the symbol
``1''). Analogously, for the series of prime numbers the events are
the appearance of the symbol ``1'', that is the fact that an integer
is prime. Hence the distance between two different events is the
distance between two consecutive prime numbers. The series of the
distances between consecutive prime numbers has been analyzed with
different complexity methods, see for example \cite{stanley},
\cite{scafetta} and references therein.

In this section we compare the behavior of the distances between
events for the prime numbers and the intermittent dynamical systems
$T_{(a,2)}$. 

For prime numbers there are many different results about the growth of
the distance $(p_{n+1} - p_{n})$ between consecutive primes. Using the
probabilistic approach, Cram\'er (\cite{cramer1}) gave the conjecture 
\begin{equation}
\limsup\limits_{n\to \infty} \ \frac{p_{n+1} - p_n}{\log^2 p_n} =1
\label{congcram}
\end{equation}

\begin{figure}[ht]
\psfig{figure=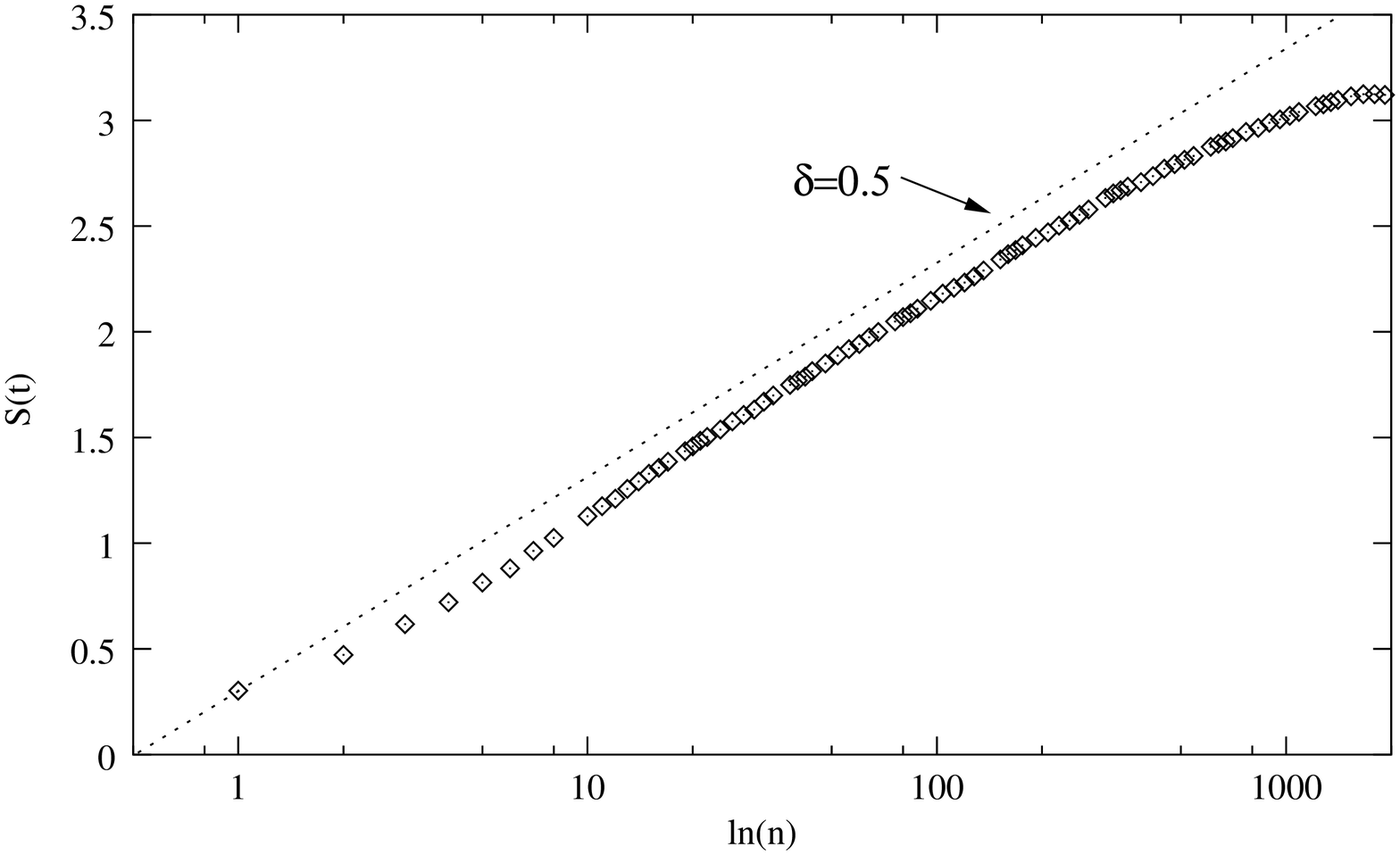,width=14cm,angle=0} 
\caption{\it EIG between the prime numbers distances and the
hypothetical trend $\log^2(p_n)$.}
\label{primivsdist}
\end{figure}

To investigate this relation, we have built the distance time series
from the prime numbers series and we have applied the EIG method. In
Figure \ref{primivsdist} we shown as once more, the hypothetical trend
for the distances between primes, given by Cram\'er conjecture, seems
to contain most part of the information, yielding a $\delta = 0.5$,
sign of independent fluctuations.

\vskip0.3cm
\vskip0.3cm

This result can be rephrased in terms of the time $t_r$ needed to have
$r$ consecutive ``0''s in $\bar \omega$. Indeed it holds that
$t_r(\bar \omega) \ge \exp (\sqrt{r})$.

The analysis of the same quantity $t_r$ was done in \cite{isola} for
the family of maps $T_{(a,2)}$. The result is that for $r$ big enough 
\begin{equation}
K^{-1} \le \frac{\E_\rho[t_r]}{r \log r} \le K
\label{teoisola}
\end{equation}
for some positive constant $K$, where the mean $\E_\rho$ is done with
respect to a probability measure $\rho$, absolutely continuous with
respect to the Lebesgue measure.

Hence if from the point of view of the information content of the
orbits, the intermittent map $T_{(0.5,2)}$ seems to have, in mean with
respect to the Lebesgue measure, the same behavior as $\bar \omega$,
the results on the distances between consecutive events show big
differences between the two different kinds of intermittency, as was
suggested by the EIG results. If we also notice that the family of
Manneville maps $T_z$ has a phase transition at $z=2$, we can conclude
that the prime numbers are well approximated by an intermittent
dynamical system similar to Manneville maps only to some extent.

\section{Conclusion}

We have shown the behavior of two complexity methods, Computable
Information Content and Entropy Information Gain, on the prime numbers
series, with an open eye on possible relations with dynamical systems,
using theoretical and numerical results.

This paper is supposed only as a tentative to open a different
approach to study prime numbers, the dynamical approach. Showing at
the same time, the importance of different complexity notions, that on
time series can show analogies, but also deep differences, as in this
case.

Moreover we remark that in our approach we looked for a deterministic
model for prime numbers, where the model is supposed to be fixed. A
different approach, similar to the spirit of the analysis in
\cite{scafetta}, is to consider a model not fixed, but with
characteristics varying with time. Our idea is however that to guess
how the characteristics of the model has to vary with time is a
problem of the same difficulty of the approximation of the function
$\pi(n)$ of the Prime Numbers Theorem.

\end{document}